\documentclass{ws-mpla}

\usepackage{amsmath}
\usepackage{verbatim}
\usepackage{amsfonts}
\usepackage{latexsym}
\usepackage{color}

\renewcommand{\d}{\mathrm{d}}

\newcommand{\Ra}{\Rightarrow}

\DeclareMathSymbol{\mg}{\mathrel}{symbols}{"1D}

\renewcommand{\gg}{\gamma}
\newcommand{\gd}{\delta}

\newcommand{\gf}{\phi}

\newcommand{\gm}{\mu}
\newcommand{\gn}{\nu}

\newcommand{\gl}{\lambda}
\newcommand{\gr}{\rho}
\newcommand{\gth}{\theta}
\newcommand{\gvth}{\vartheta}
\newcommand{\gs}{\sigma}
\newcommand{\gt}{\tau}

\newcommand{\gp}{\pi}
\newcommand{\gps}{\psi}
\newcommand{\get}{\eta}

\newcommand{\gD}{\Delta}

\newcommand{\gL}{\Lambda}

\newcommand{\gTh}{\Theta}

\newcommand{\cD}{{\cal D}}
\newcommand{\cE}{{\cal E}}
\newcommand{\cF}{{\cal F}}
\newcommand{\cG}{{\cal G}}

\newcommand{\cT}{{\cal T}}

\newcommand{\ui}{{\underline i}}
\newcommand{\uj}{{\underline j}}

\newcommand{\tr}{\text{tr}}

\newcommand{\ra}{\rightarrow}

\newcommand{\inv}{^{-1}}
\newcommand{\dsp}{\displaystyle}

\newcommand{\labl}[1]{\label{#1}}
\newcommand{\beq}{\begin{equation}}
\newcommand{\eeq}{\end{equation}}
\newcommand{\barr}{\begin{array}}
\newcommand{\earr}{\end{array}}
\newcommand{\equ}[1]{\begin{gather} #1 \end{gather}}
\newcommand{\equa}[1]{\begin{align} #1 \end{align}}

\newcommand{\tabu}[2]{\begin{tabular}{#1} #2 \end{tabular}}
\newcommand{\arry}[2]{\begin{array}{#1} #2 \end{array}}

\newcommand{\non}{\nonumber}
\newcounter{oldcounter}

\newcommand{\bder}{\bar\partial}
\newcommand{\bgl}{{\bar\lambda}}

\newcommand{\bgt}{{\bar\tau}}

\newcommand{\tgD}{{\tilde \Delta}}

\newcommand{\Intr}{\mathbb{Z}}
\newcommand{\Cplx}{\mathbb{C}}
\newcommand{\Real}{\mathbb{R}}

\newcommand{\ba}[2]{\[\begin{array}{#2}\label{#1}}
\newcommand{\ea}{\end{array}\]}
\newcommand{\be}{\begin{equation}}
\newcommand{\ee}{\end{equation}}
\newcommand{\bea}{\begin{eqnarray}}
\newcommand{\eea}{\end{eqnarray}}

\newcommand{\E}[1]{\mathrm{E_{#1}}}
\newcommand{\U}[1]{\mathrm{U(#1)}}
\newcommand{\SU}[1]{\mathrm{SU(#1)}}
\newcommand{\SO}[1]{\mathrm{SO(#1)}}

\newcommand{\brkt}[2]{\bigl[ ^{#1}_{#2} \bigr]}

\newcommand{\rep}[1]{\mathbf{#1}}
\newcommand{\crep}[1]{\overline{\rep{#1}}}

\begin{document}

\markboth{S.\ Groot Nibbelink}
{Shape of gauge field tadpoles in string theory}

%
\catchline{}{}{}{}{}
%

\title{Shape of gauge field tadpoles in heterotic string theory}

\author{Stefan Groot Nibbelink}

\address{William I.\ Fine Theoretical Physics Institute,
University of Minnesota, \\ 
Minneapolis, MN 55455, USA, 
nibbelin@physics.umn.edu}

\maketitle

\begin{abstract}

Orbifolds in field theory are potentially singular objects for at
their fixed points the curvature becomes infinite, therefore one may
wonder whether field theory calculations near orbifold singularities
can be trusted. String theory is perfectly well defined on orbifolds
and can therefore be taken as a UV completion of field theory on
orbifolds. We investigate the properties of field and string theory
near orbifold singularities by reviewing the computation of a one loop
gauge field tadpole. We find that in string theory the twisted states
give contributions that have a spread of a couple of string lengths
around the singularity, but otherwise the field theory picture is
confirmed. One additional surprise is that in some orbifold models one
can identify local tachyons that give contributions near the orbifold
fixed point.

\end{abstract}

\ccode{PACS Nos.: 11.10.Kk,11.25.Wx}

\section{Field theory on orbifolds}
\labl{sc:orbiField}

In recent years there has been an enormous effort to understand the
physics of extra dimensions. To study consequences of extra dimensions
in all generality is a formidable task, therefore most fruitful
investigations have relied on simple ans\"atze because they tell us
what kind of properties we need in the extra dimensions in order to
construct realistic models of particle physics. 

The simplest choice for the extra dimensions is to have just only one
extra dimension that has the topology of a circle of radius $R$. We
parametrize the circle by the periodic coordinate: $y \sim y + 2\gp\, R$. 
From an effective four dimensional point of view this five dimensional
theory contains an infinite number of fields, since one can perform
the following Kaluza--Klein decomposition
\equ{
\gf(x,y) = \sum_{{n} \in \Intr} 
\gf_n(x) {e^{i n y/ R } }, 
~~ \text{and} ~~
x \in \Real^4
\labl{KKexp} 
}
of a field $\gf$ in five dimensions. 
But if fermions live in the bulk of the extra dimensions, we face 
immediately a fundamental problem. Again let us illustrate the
point with the a single extra dimension compactified on a
circle. Since $\gg_5$ is part of the five dimensional Clifford
algebra, fermions in five dimensions cannot be chiral. Therefore the
modes $\gf_n(x)$ in the Kaluza--Klein expansion \eqref{KKexp} are
Dirac fermions that do not have a definite chirality. But we know that
the electroweak interactions do not treat left-- and right--hand
particles equally. This shows that the simplest compactification on a
circle can never lead to a semi--realistic fermion spectrum if the
fermions live in the bulk.

\begin{figure}[th]
\[
\epsfig{file=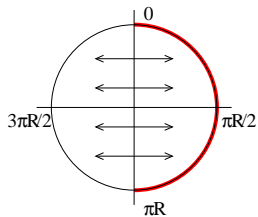, width=3.3cm, angle =0}
\]
\caption{The orbifold $S^1/\Intr_2$ is defined as the circle modded by
a discrete $\Intr_2$ symmetry, the resulting space is a line segment. 
\labl{fg:orbiS1Z2}}
\end{figure}

To overcome this hurtle one needs to do something drastic because the
properties of the chiral spectrum in four dimensions are determined by
the number of zero of the Dirac operator in the extra dimensions. As
the number of zero modes is a topological number, it is impossible to
alter it by making some continuous changes of parameters. This is the
place where orbifolds may come to our rescue. Before addressing how
the number of zero modes change on an orbifold, let us first expose
some basic properties of the simplest orbifold $S^1/\Intr_2$. Consider
again the circle, but assume that there is now also a reflection
symmetry acting of the space $y \ra - y$. The orbifold is depicted in
figure \ref{fg:orbiS1Z2}. Then for each (bosonic) field one has to
make a choice whether it is even or odd under this parity. Only an
even field has a zero mode from the four dimensional perspective. 
For a fermion the situation is slightly more involved: in
order for the $\Intr_2$ action to be a symmetric of the fermionic
kinetic action it has to act as 
\equ{
\gps(-y) = \gg_5 \gps(y). 
} 
This implies that the four dimensional left-- and right--handed chiral
Kaluza--Klein modes have are even and odd, respectively. Moreover only
the left--handed Kaluza--Klein fermions have a zero mode. This shows
that the problem of obtaining chiral spectra from extra dimensions can
be evaded by using orbifolds.

But clear we have paid a price: The orbifold $S^1/\Intr_2$ is a
singular manifold because at the points $0$ and $\pi R$ it is not
differentiable. Therefore, one might wonder whether a field theory on
an orbifold beyond just being an interesting mathematical construction,
is physically well--defined. Field theories in extra dimensions are
generically non--renormalizable, this means that they can only be
understood as effective theory valid up to a cut--off scale
$\gL$. This means that on distances smaller than the inverse cut--off
we cannot trust our field theoretical calculations. We will
investigate this issue in detail in the next section where we compute
the expectation value of a specific operator on an orbifold in field
theory and show that the results are indeed rather singular in a
variety of ways. Then we get to the main objective of this paper is to
investigate how string theory can regulates the potential problems
that field theory has near orbifold singularities. We have chosen to
attack this question by considering gauge field tadpoles in both field
and string theory.

\section{The gauge field tadpole}
\labl{sc:fieldtad}

In this section we review the results from field theoretical
calculations of gauge field tadpoles on orbifolds. This is interesting
in its own right as well as becomes clear when we put our work in
context.  

The local gauge field tadpoles are closely related to the well--known
tadpoles for the $D$--terms \cite{Fayet:1974jb}  in $N=1$
supersymmetric gauge theories. Such tadpoles can only arise at
one--loop provided that the sum of charges of chiral matter multiplets
is non--vanishing \cite{Fischler:1981zk}. Such Fayet--Iliopoulos
$D$--terms have also been investigated in heterotic
\cite{Dine:1987xk,Atick:1987gy,Dine:1987gj} and type I string
\cite{Poppitz:1998dj} context. Four dimensional Fayet--Iliopoulos
$D$--terms have a natural generalization to five dimensional
supersymmetric gauge theories on orbifolds. As observed in 
ref.\ \cite{Mirabelli:1998aj} the effective auxiliary field at the
boundaries of $S^1/\Intr_2$ also contains a derivative with respect to
the fifth dimension of a real scalar field part of the five dimensional
gauge multiplet. There have been several calculations showing that
such interactions are generated at the one--loop level
\cite{Ghilencea:2001bw,Barbieri:2001cz,Scrucca:2001eb,GrootNibbelink:2002wv,GrootNibbelink:2002qp}.
For bulk states the result
is given by a massive quadratically divergent integral times
delta--functions at the orbifold singularities. Here the ``mass''
represents the derivative squared  with respect to the fifth dimension
acting on  these delta--functions. By investigating the required
counter--term structure for this real scalar field of the gauge
multiplet, it has been argued in refs.\ 
\cite{GrootNibbelink:2002wv,GrootNibbelink:2002qp,Lee:2003mc} 
that these tadpoles may lead to strong localization of bulk zero
modes. This treatment has not been universally accepted, for
alternatives see refs.\  \cite{Barbieri:2002ic,Marti:2002ar}. 
The existence of Fayet--Iliopoulos tadpoles for both auxiliary and
derivative of physical fields is in no way particular to five
dimensional models. One--loop computations in higher dimensional field
theoretical models  have confirmed that these tadpoles are generated
locally at orbifold fixed points \cite{vonGersdorff:2002us,Csaki:2002ur} .
As the main objective of this paper is to compare field and string
theory calculations on orbifolds with each other, we review here the
tadpole computation  in field theory approximation of heterotic string
models on $\Cplx^3/\Intr_3$ \cite{GrootNibbelink:2003gb}.

\begin{figure}[th]
\[
\raisebox{0ex}{\scalebox{.7}{\mbox{\input{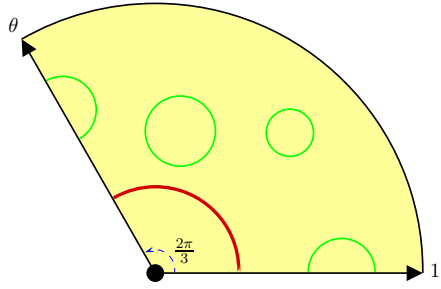}}}}
\]
\caption{
The fundamental domain of the orbifold $\Cplx^3/\Intr_3$ has the shape
of a six dimensional cone. Here we have drawn the its two dimensional
analog $\Cplx/\Intr_3$. By identifying the edges one obtains a cone
with a deficit angle of $2 \gp/3$. 
\labl{fg:orbiC3Z3}}
\end{figure}

The orbifold $\Cplx^3/\Intr_3$ defined as the three
dimensional complex plane $(z_1, z_2, z_3) \in \Cplx^3$ on which the
diagonal orbifold twist 
\equ{ 
(z_1, z_2, z_3) \ra (\gth z_1, \gth z_2, \gth z_3), 
\qquad 
\gth = e^{2\pi i/3}, 
\labl{orbitw}
} 
acts. This orbifold has one singularity at the origin of real
codimension six. In figure \ref{fg:orbiC3Z3} we have drawn the
fundamental domain of the two dimensional analog, $\Cplx/\Intr_3$ or
the orbifold $\Cplx^3/\Intr_3$. The low energy spectrum of the
heterotic string theory contains a ten dimensional Yang--Mills
theory. In fact there are two heterotic string theories, we focus here
on the one where gauge group of the Yang--Mills theory is $\E{8}\times
\E{8}$. The action of the orbifold twist \eqref{orbitw} is extended to
the gauge bundle by the action 
\equ{
A(x, z) \ra A(x, \gTh\, z) = U\, A(x, z)\, U\inv,
\qquad 
U = e^{2\gp i\, v_a^I H_{a}^I}, 
} 
on the gauge field one--form $A$. The generators $H^I_1$ and $H^I_2$
of the Cartan subgroups of  $\SO{16} \subset \E{8}$  
and $\SO{16}' \subset \E{8}'$ are used to define the twist action as
shifts of the lattices spanned by the $\SO{16} \times \SO{16}'$ roots
and spinorial weights. The gauge shift vectors $v_a = (v_a^I)$ are 
quantized in multiples of $\frac 13$. These gauge shifts determine the
full spectrum of the theory. The untwisted chiral matter
representation $\rep{R}$ has a three fold degeneracy. In addition
absence of local gauge and mixed gravitational anomalies requires the
introduction of four dimensional states at the fixed point. Contrary
to field theory these states are completely determined in the twisted
sectors of the heterotic string as we explain the next section. 
The twisted states that come in three equal copies are denoted by 
$\rep{T}$, while others are referred as $\rep{S}$. The five possible
spectra at the fixed point are given in table \ref{tb:z3models}.

\begin{table}[ht]
  \tbl{The spectra at the fixed point of the five $\Intr_3$ orbifold models
are displayed.    \label{tb:z3models}}  
{\tiny
\renewcommand{\arraystretch}{1.25}
  \begin{tabular}{|l|l|l|ll|l}\hline
    {Model}
      & {Shift $(v_1^I ~|~ v_2^I)$ and }
      & {Untwisted}
      & {Twisted}
      & 
\\
& {gauge group $G $} 
& {$(\rep{3}_H,\rep{R})$ }
& {$(\rep{1}_H, \rep{S})$ } 
& {$(\rep{\bar{3}}_H, \rep{T})$}
      \\\hline
    $\E{8}$
      & $\frac{1}{3}\!\left(~0^8 ~~~~ ~~~~~ ~~|~~ 0^8 ~~~~~ ~~~~~ \right)$
      & 
      &  
      & $(\rep{1})(\rep{1})'$
      \\
      &    $~~~~~~ ~~~~~~ ~\, \E{8}  \times   \E{8}'$
      & & & 
      \\\hline
    $\E{6}$
      & $\frac{1}{3}\!\left(\mbox{-} 2,~1^2,~ 0^5 ~~|~~ 0^8 ~~~~~ ~~~~~ \right)$
      & $(\rep{27},\crep{3})(\rep{1})'$
      & $(\rep{27},\rep{1})(\rep{1})'$
      & $ (\rep{1},\rep{3})(1)'$
      \\
       & $~~~ \E{6}\!\times\!\SU{3} \times \E{8}' ~~~~~~~ $
      & & &
      \\\hline
    $\E{6}^2$
      & $\frac{1}{3}\!\left(\mbox{-}2,~1^2,~0^5 ~~|~~ \mbox{-}2,~1^2,~0^5\right)$
      & $(\rep{27},\crep{3})(\rep{1},\rep{1}) \!+\! (\rep{1},\rep{1})(\rep{27},\crep{3})'$
      & $(\rep{1},\rep{3})(\rep{1},\rep{3})'$
      &  
      \\
      & $~~~ \E{6}\!\times\!\SU{3} \times \E{6}'\!\times\!\SU{3}'$
      &  & & 
      \\\hline
    $\E{7}$
      & $\frac{1}{3}\!\left(~0,~1^2,~0^5 ~~|~~ \mbox{-}2,~0^7 ~~~~~ \right)$
      & $(\rep{1})_{0}(\rep{64})'_{\frac{1}{2}}
 + (\rep{56})_{1}(\rep{1})'_{0}$
      & $(\rep{1})_{\frac{2}{3}}(\rep{14})'_{\mbox{-}\frac{1}{3}} $
      & $ (\rep{1})_{\frac{2}{3}}(\rep{1})'_{\frac{2}{3}}$
      \\
      &  $ ~~~~\,  \E{7}\!\times\!\U{1} \times
\SO{14}'\! \times\! \U{1}' \!\!$
      & $+\, (\rep{1})_{0}(\rep{14})'_{\mbox{-}1} + (\rep{1})_{\mbox{-}2}(\rep{1})'_{0}$
      & $ +\,  (\rep{1})_{\mbox{-}\frac{4}{3}}(\rep{1})'_{\frac{2}{3}}$ 
      & 
      \\\hline
    $\SU{9}$
     & $ \frac{1}{3}\! \left(\mbox{-}2,~1^4~,0^3 ~~|~~ \mbox{-}2,~0^7
~~~~~ \right) $
     & $(\rep{84})(\rep{1})'_{0} + (\rep{1})(\rep{64})'_{\frac{1}{2}}$
      & $(\crep{9})(\rep{1})'_{\frac{2}{3}}$
     & 
      \\
      & $~~~~~ ~~~~ \SU{9} \times \SO{14}'\!\times\!\U{1}' \!\!$
      &$+\, (\rep{1})(\rep{14})'_{\mbox{-}1}$ & &
      \\\hline
    \end{tabular}
 }
\end{table}

Only two of these models contain $\U{1}$ factors which can be
anomalous. The result of the calculation of the gauge field tadpole on
the orbifold $\Cplx^3/\Intr_3$ can be cast in the form
\cite{GrootNibbelink:2003gb,GNL_I} 
\equ{
\raisebox{-3mm}{
\epsfig{file=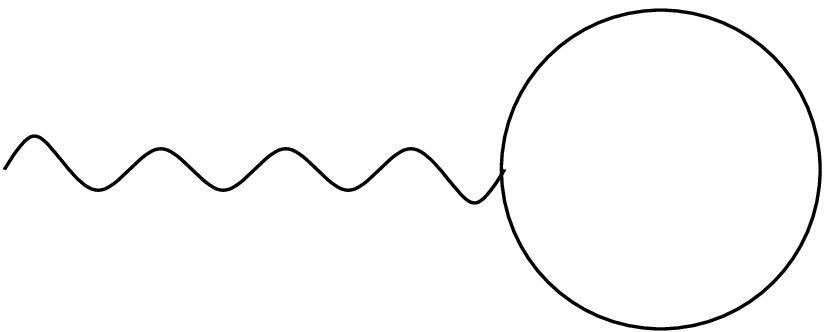,width=18mm}}
= 
\langle F_{j\uj}^b (k) \rangle 
=
\frac{\gd^4(k_4)}{(2\gp)^4} \, 
\frac {\gp}4 \gL^2 \, 
\int\limits_{\mbox{-}\frac 12}^{\frac 12} \d \gt_1
\int\limits_{1}^\infty \frac {\d \gt_2}{\gt_2^2} \, 
\sum_{s = un, tw}\, Q_s^b \, 
e^{- \gD_s\, k_i k_\ui /\gL^2}
.
\labl{TadFieldSchwinger}
}
The quantities $\smash{Q_{un}^b, Q_{tw}^b, \gD_{un}}$ and
$\smash{\gD_{tw}}$ appearing in this expression are given by 
\equ{
Q_{un}^b = \frac {3}{27}\,  \tr_{\rep{R}}(q_b),
~~ 
Q_{tw}^b =  \tr_{\rep{S}}(q_b) + 3 \,\tr_{\rep{T}}(q_b), 
\quad  
\gD_{un} = 4\gp\, \gt_2\, \frac 13, 
~~  
\gD_{tw} =0. 
\labl{QgDfield}
}
The subscripts $un$ and $tw$ refer respectively to the untwisted and 
twisted sectors. Notice that the $\gt_1$ integration is redundant, 
it has been included for the subsequent comparison with the string
result. The $\gt_2$ integral arises by using a variant of Schwinger's
proper time regularization \cite{Polchinski:1986zf} to rewrite the
momentum integral of the scalar propagator 
\equ{
\int \frac{\d^4 p_4}{p_4^2 + m^2} = 
 \frac {\gp}4 \gL^2\, \int_{-\frac 12}^{\frac 12} \d \gt_1 \, 
\int_{1}^\infty \frac{\d \gt_2}{\gt_2^2}\,  
e^{-4\gp\, \gt_2\, m^2/\gL^2},
\labl{Schwinger}
}
for an arbitrary mass $m$ and a cut--off scale $\gL$.

\section{String theory on orbifolds} 
\labl{sc:orbiString}

To date we have essentially only a single proposal for an
ultra--violet complete theory: String theory. In this work we restrict
ourselves to the well--studied heterotic closed string theory. It has
been shown that all amplitudes in this theory are finite and therefore
well--defined. Therefore one may hope that it is possible to 'lift'
field theories of extra dimensions in string theory, which 
leaves plenty of room for extra dimensions since it target space needs
to be ten dimensional for internal consistency. Hence for  string
theory to make contact with our four dimensional world six dimensions
have to be compactified. The issue of chiral fermions in four
dimensions reappears and one has to choose a specific space for the
compactification that allows for a chiral spectrum. Moreover, all
consistent string theories are supersymmetric in ten dimensions, which
means that a simple toroidal compactification of six dimensions leads
to a theory with at least $N=4$ supersymmetry  in four dimensions. It
has been shown that by compactifying on special manifolds called
Calabi--Yau the $N=4$ supersymmetry is broken down to $N=1$
supersymmetry in four dimensions. These smooth Calabi--Yau manifolds
are rather complicated objects, but fortunately many of their crucial
properties, like breaking sufficient amounts of supersymmetry, are
shared by suitable chosen orbifolds.

There are many orbifolds known that have the appropriate properties to
be interesting in principle for four dimensional string
phenomenology. However, the main focus of this review is not so much
on possible phenomenological applications, but on the investigation
what happens near an orbifold singularity in both field theory and
string theory. Therefore, the requirement that in four dimensions we
have a finite Planck mass is not a pressing issue for our investigation. 
This allows us to perform our investigation on the non--compact
orbifold $\Cplx^3/\Intr_3$ which has a simpler structure, see figure 
\ref{fg:orbiC3Z3}.  We have depicted two types of closed
string: untwisted and twisted strings in that figure. 
The latter winds around the orbifold singularity at the top of the
cone. These twisted strings are special in that they are localized
around the singularity, and for that reason one usually assumes that
they can be represented as fields living exactly at the singularity in
field theory.  Their spectrum is given in table \ref{tb:z3models}.

Contrary to field theories, string theories are perfectly
well--defined on orbifolds. The reason for this is simple: In string
theory, what we call coordinates $z_i$ in field theory, have become
fields $X_i(\gs)$ on the two dimensional string world sheet
parameterized by the complex variable $\gs$. For a closed string there
is always at least one cyclic direction on this world sheet. The
$\Intr_3$ orbifolding described geometrically above is implemented by
having different boundary conditions when one goes around closed
cycles of the string world. For example for a tree diagram in string
theory, 
\equ{
\raisebox{-2ex}{\scalebox{0.8}{\mbox{\begin{picture}(0,0)%
\includegraphics{cilinder.pstex}%
\end{picture}%
\setlength{\unitlength}{3947sp}%
\begingroup\makeatletter\ifx\SetFigFont\undefined%
\gdef\SetFigFont#1#2#3#4#5{%
  \reset@font\fontsize{#1}{#2pt}%
  \fontfamily{#3}\fontseries{#4}\fontshape{#5}%
  \selectfont}%
\fi\endgroup%
\begin{picture}(1369,479)(2768,-5625)
\put(3151,-5461){\makebox(0,0)[lb]{\smash{\SetFigFont{12}{14.4}{\rmdefault}{\mddefault}{\updefault}{\color[rgb]{.82,0,0}$\gs$}%
}}}
\end{picture}
}}}
\qquad 
X_i({\gs + 1}) = {\gth^p} \, X_i({\gs}),
\quad \gth = e^{2 \pi i/3} 
}
there are three different boundary conditions, labeled by $p = 0,1,2$. 
These boundary conditions define different orbifold sectors: 
The untwisted sector satisfies the trivial boundary condition with 
$p = 0$. The twisted sectors $p =1,2$ have the center of mass at one
of the fixed point: $ \langle X_i \rangle = 0$. Clearly these boundary
conditions on the string world sheet for both the untwisted and
twisted states are entirely smooth, even though the center of mass of
the twisted states is localized at the orbifold singularity.

The coordinate fields of closed strings can be left-- or
right--moving. The heterotic string does not only contain the
coordinate fields $X^M(\gs)$ but also world sheet fermions. The
right--moving fermions $\gps^M(\gs)$ are the world sheet super
partners of the right--moving coordinate fields. The left--moving
fermions $\gl^I_a(\gs)$ are not the super partners of the left--moving
coordinate fields. These fermions generate the $\E{8}\times \E{8}$
gauge symmetry: The index $a$ labels the two $\E{8}$'s and $I$ the
Cartan elements within $\E{8}$ algebra. With these ingredients we can 
move to the string computation of the gauge field tadpole.

\section{Stringy computation of the gauge field tadpole} 
\labl{sc:stringtad}

\begin{figure}
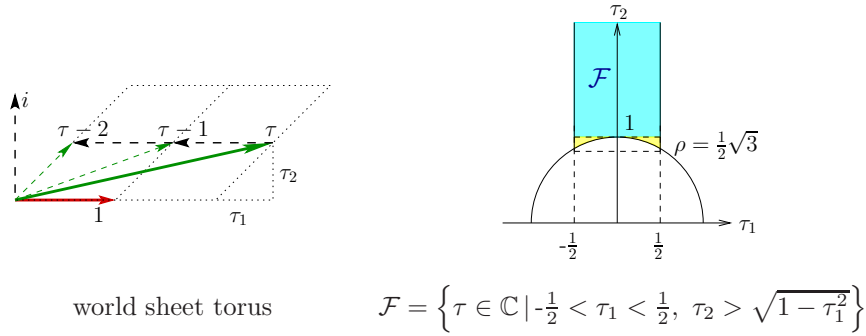

\[
\arry{ccc}{
\raisebox{2ex}{\scalebox{0.6}{\mbox{\input{torus.pstex_t}}}}
&  \quad & 
\raisebox{0ex}{\scalebox{0.6}{\mbox{\input{FunDom.pstex_t}}}}
\\[2ex] 
\text{world sheet torus} && 
\cF = \Bigl\lbrace 
\gt \in \Cplx \, | \, \mbox{-}\frac 12 < \gt_1 < \frac 12, ~
\gt_2 > \sqrt{ 1 - \gt_1^2 }
\Bigr\rbrace
}
\]
\caption{
The first picture gives a torus defined by the complex variable 
$\gt$. To label inequivalent tori this parameter is restricted to lie
in the fundamental domain $\cF$, depicted in the second picture. 
\labl{fg:ModularPara}
}
\end{figure}

We now turn to the calculation of the gauge field tadpole in string
theory. As the heterotic theory only contains closed strings, the
string diagram takes the form 
\equ{
\arry{ccc}{
\raisebox{-3.5ex}{\scalebox{.9}{\mbox{\begin{picture}(0,0)%
\includegraphics{StringTad.pstex}%
\end{picture}%
\setlength{\unitlength}{2881sp}%
\begingroup\makeatletter\ifx\SetFigFont\undefined%
\gdef\SetFigFont#1#2#3#4#5{%
  \reset@font\fontsize{#1}{#2pt}%
  \fontfamily{#3}\fontseries{#4}\fontshape{#5}%
  \selectfont}%
\fi\endgroup%
\begin{picture}(3097,902)(2135,-1712)
\end{picture}
}}}
&
\Ra 
& 
\raisebox{-3.5ex}{\scalebox{.9}{\mbox{\begin{picture}(0,0)%
\includegraphics{StringTadVer.pstex}%
\end{picture}%
\setlength{\unitlength}{2881sp}%
\begingroup\makeatletter\ifx\SetFigFont\undefined%
\gdef\SetFigFont#1#2#3#4#5{%
  \reset@font\fontsize{#1}{#2pt}%
  \fontfamily{#3}\fontseries{#4}\fontshape{#5}%
  \selectfont}%
\fi\endgroup%
\begin{picture}(2790,914)(2394,-1718)
\end{picture}
}}}
}
\non 
}
By exploiting the conformal symmetry of the string theory, the
external leg of the string diagram can be squeezed to a line ending at
a vertex operator on the torus. The vertex operator corresponding to
an internal gauge field in the Cartan subalgebra  is expressed in
terms of the world sheet fields as 
\equ{
V_j^{bJ} =\
: \! ( \bder X_j + i \, k_M \gps^M \gps_j) \,
\bgl^{J}_b \gl^{J}_b \, e^{i k_M X^M} \! : 
\labl{vertexoperator}
}
Since the vertex operator leads to the evaluation of powers sheet
fields at the same point of the string world sheet, this expression is
ill--defined unless it is normal ordered, which is indicated by 
the $: \ldots :$ notation. Schematically the evaluation of this vertex
operator on the world sheet torus takes the form of the path integral
average 
\equ{
{\langle V_j^{bJ} \rangle} = 
\sum_{\text{sectors}}\  \int_{{\cF}} \frac{\d^2 \gt}{\gt_2^2} \ 
{\int\! \cD X \cD \gps \cD \gl \ }
e^{-S_{\text{free}}} \ {V_j^{bJ}}. 
\labl{stringTad}
}
This expression requires some further explanation: The one--loop
diagram of the closed string is a torus. Conformally inequivalent tori
are labeled by a complex modulus $\gt$ that lies with the fundamental
domain $\cF$ of the modular group depicted in figure
\ref{fg:ModularPara}. The shape of this fundamental domain motivated
the use of the Schwinger proper time regularization employed in 
\eqref{TadFieldSchwinger}. This regularization mimics the string range
of integration quite well, only the light (yellow) shaded region are
ignored. And finally the sum over ``sectors'' all possible orbifold
boundary conditions. As the one--loop string diagram is a torus, which
as two non--contractible cycles, there are in total nine different
boundary conditions:  
\equ{
\arry{ccc}{ 
\hspace{-2ex} 
\raisebox{-7ex}{\scalebox{1.8}{\mbox{\begin{picture}(0,0)%
\includegraphics{Torus.pstex}%
\end{picture}%
\setlength{\unitlength}{2881sp}%
\begingroup\makeatletter\ifx\SetFigFont\undefined%
\gdef\SetFigFont#1#2#3#4#5{%
  \reset@font\fontsize{#1}{#2pt}%
  \fontfamily{#3}\fontseries{#4}\fontshape{#5}%
  \selectfont}%
\fi\endgroup%
\begin{picture}(1390,937)(3807,-1735)
\end{picture}
}}}
& \qquad & 
\arry{l}{
X^\gm (\gs +1 ) = X^\gm(\gs + \gt) = X^\gm(\gs), 
\\[2ex]
X^i({\gs + 1}) = {\gth^{p\,}} \, X^i({\gs}),
\\[2ex] 
X^i({\gs + \gt}) = {\gth^{p'}} \, X^i({\gs}),
~~ 
\gth = e^{2\pi i/3},
}
}
\labl{sectors} 
}
labeled by $p, p' = 0,1,2$. In table \ref{tb:orbiclass} we have given
a classification of these boundary conditions. Notice that this is finer
than the standard classification of ten dimensional, four dimensional
untwisted and twisted states. The reason for this will become
apparent below.

\begin{table}[ht]
  \tbl{Classification of boundary conditions.\labl{tb:orbiclass}}  
{\tabu{| r l | l | l  | }{
\hline 
dim & description & $s$  & $(p, p')$  
\\[0ex] \hline &&  &  \\[-2ex]
10D & states & & $(0,0) $
\\[1ex]
10D &  untwisted & $u$ & $(0, 1); ~~ (0, 2) $
\\[1ex]
4D & twisted & $t$ & $(1,0); ~~(2, 0)$
\\[1ex] 
4D & double twisted & $d_+$ & $(1,1); ~~ (2,2)$ 
\\ 
& & $d_-$ & $(1,2); ~~ (2,1)$
\\[0ex]
\hline 
}}
\end{table}

In refs.\ \cite{GNL_I} a detailed account is given of the
computation of the tadpole \eqref{stringTad} using the fact that the
world sheet theories for $X^M$, $\gps^M$ and $\gl^I_a$ are free for
orbifolds. The resulting shape can be cast in the form of Gaussian
distributions  
\equ{
G^{bJ}(k) =
\int_\cF \frac{\d^2 \gt}{\gt_2^2} 
\sum_{s=(p,p')\neq 0}
Q^{bJ}_s(\gt) 
\, 
e^{- \gD(\gt) \,k^\gm k_\gm - \gD_X\brkt{p\, /3}{p'/3}(\gt)\, k_\ui k_i }
\labl{GaussMom}
}
convoluted by the integration over the fundamental domain $\cF$. 
The explicit form of the charge functions $Q^{bJ}_s(\gt)$ can be found
in refs.\ \cite{GNL_I,GNL_II}. The only difference between the
expressions quoted these references is that here we have also included
a Gaussian involving the four dimensional momenta $k^\gm$ with a width
set by $\gD(\gt)$.

The functions $\gD(\gt)$ and $\gD_X\brkt{p\, /3}{p'/3}(\gt)$ arise
because of the normal ordering in the definition of the vertex operator
\eqref{vertexoperator} to avoid singular products of coordinate fields
$X^M$. More precisely, the propagator for the untwisted world sheet
bosons $X^\gm(\gs)$ reads 
\equ{
\langle X^\gm(\gs) X^\gn(0) \rangle = \get^{\gm\gn} \tgD(\gs|\gt), 
\quad 
\tgD(\gs|\gt) 
= - \ln \Big[ 
2\gt_2\, 
e^{-2\pi\, \frac {\gs_2^2}{\gt_2}} \, 
\Bigl| \frac{\gvth_1(\gs| \gt)}{\gvth_1'(0|\gt)} \Bigr|^2 \Big], 
\labl{UnProp}
} 
while the twisted propagators 
\(
\tgD_X\brkt{p/3}{p'\!/3}(\gs|\gt) 
\)
can be expressed as sum of untwisted ones 
\equ{
\tgD_X\brkt{p/3}{p'\!/3}(\gs|\gt) = 
\sum_{k, k'  = 0}^2 e^{- 2\pi i(p k - p' k')/3}\, 
\tgD \Bigl( \frac {\gs + k +k' \gt }3 \Big| \gt  \Bigr).
\labl{TwProp}
}
Both the untwisted and twisted propagators develop a 
$\ln |\gs|^2$ singularity in the zero separation limit. The conformal
normal ordered propagators $\gD$ and $\gD_s$ are obtained by dropping
this singular term: 
\equa{ \dsp &
\gD(\gt) = - \ln (2 \gt_2) + \tilde c, 
\quad 
\gD_X\brkt{p/3}{p'\!/3}(\gt)  =  \gD(\gt) + \gD_s(\gt),
\quad \text{with}  
\non 
\\[2ex] \dsp &
\gD_s(\gt)  =   
2 \ln 3 + 
\sum_{(k, k') \neq 0} e^{- 2\pi i(p k - p' k')/3}\, 
\tgD \Bigl( \frac {k +k' \gt}3 \Big| \gt \Bigr).
\labl{NormTwProp}
}
with $s = (p, p')\neq 0$. 
Here $\tilde c$ denotes an arbitrary normal ordering constant which is
left undetermined when subtracting the $\ln|\gs|^2$ singularity. The
$2 \ln 3$ term for the normal order twisted propagator arises because
the singular term of the twisted correlator is $\ln|\gs/3|^2$. With
these definitions the profile becomes 
\equ{
G^{bJ}(k) =
\int_\cF \frac{\d^2 \gt}{\gt_2^2} 
\sum_{s=u,t,d_\pm}
Q^{bJ}_s(\gt) 
\, 
e^{- \gD(\gt) (k^\gm k_\gm + k_\ui k_i ) } 
\, e^{- \gD_s(\gt) k_\ui k_i }. 
\labl{GaussMom2}
}
Strictly speaking perturbative string amplitudes are only defined
on--shell, i.e.\ when $k^\gm k_\gm + k_\ui k_i =0$. By taking this
amplitude on--shell we see that the Gaussian with $\gD(\gt)$
containing the overall normal ordering constant is dropped. 
\footnote{We are indebted to E.\ Kiritsis for emphasizing that by going
on--shell without requiring momentum conservation results in less 
dependence on normal ordering constants. 
See ref.\ \cite{Antoniadis:2002cs}.} However, because Lorentz and
rotational invariance between the four dimensional and internal spaces
is lost, there is no reason why the normal ordering constants for both
spaces need to be equal. Notice that we have not enforce four
dimensional momentum conservation, $k_\gm =0$, otherwise the whole
internal momentum dependence is lost.

\section{Profile of the tadpole in string theory}

\begin{figure}
\begin{center}
\tabu{ccc}{
\includegraphics[width=3.7cm,height=3.2cm]{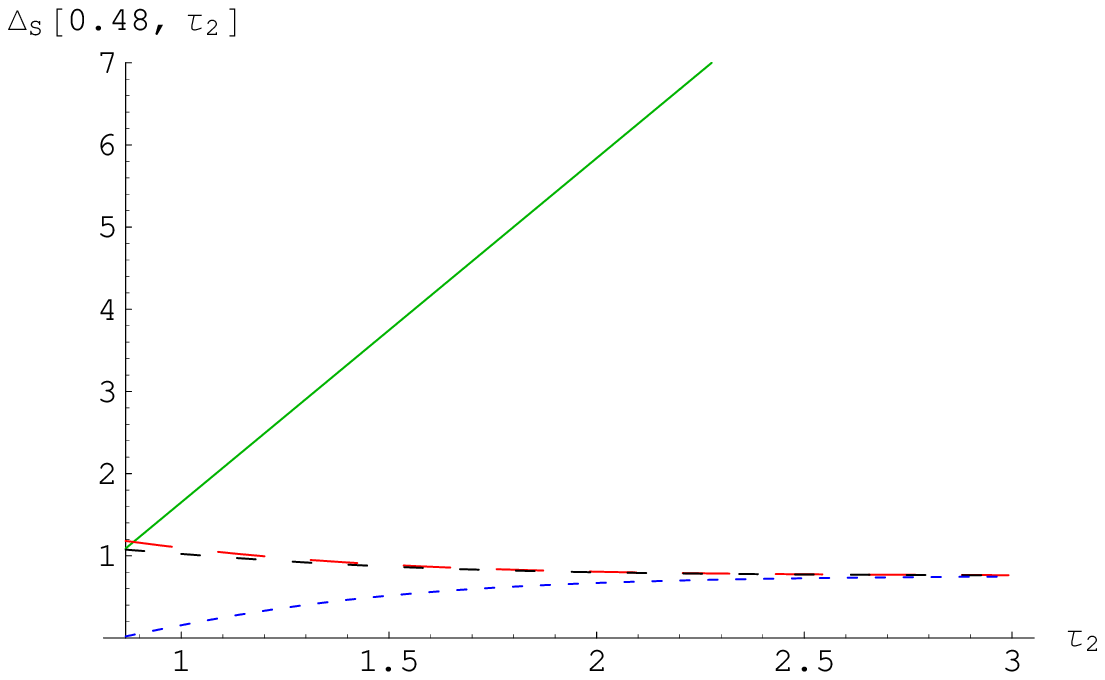} &
\includegraphics[width=3.7cm,height=3.2cm]{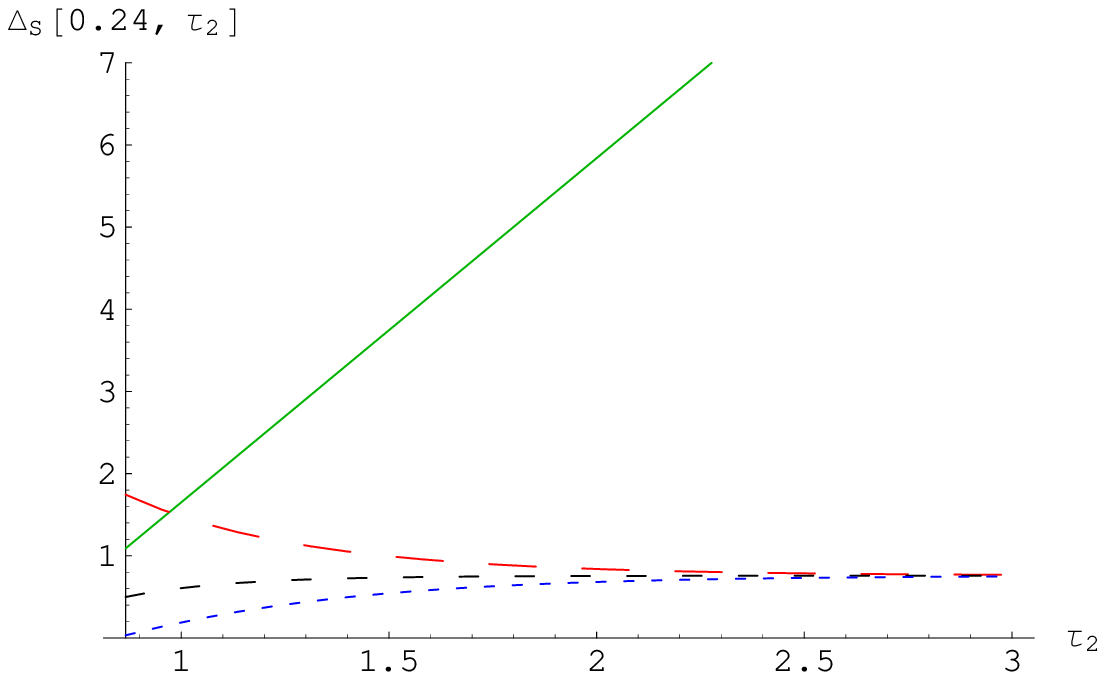} &
\includegraphics[width=3.7cm,height=3.2cm]{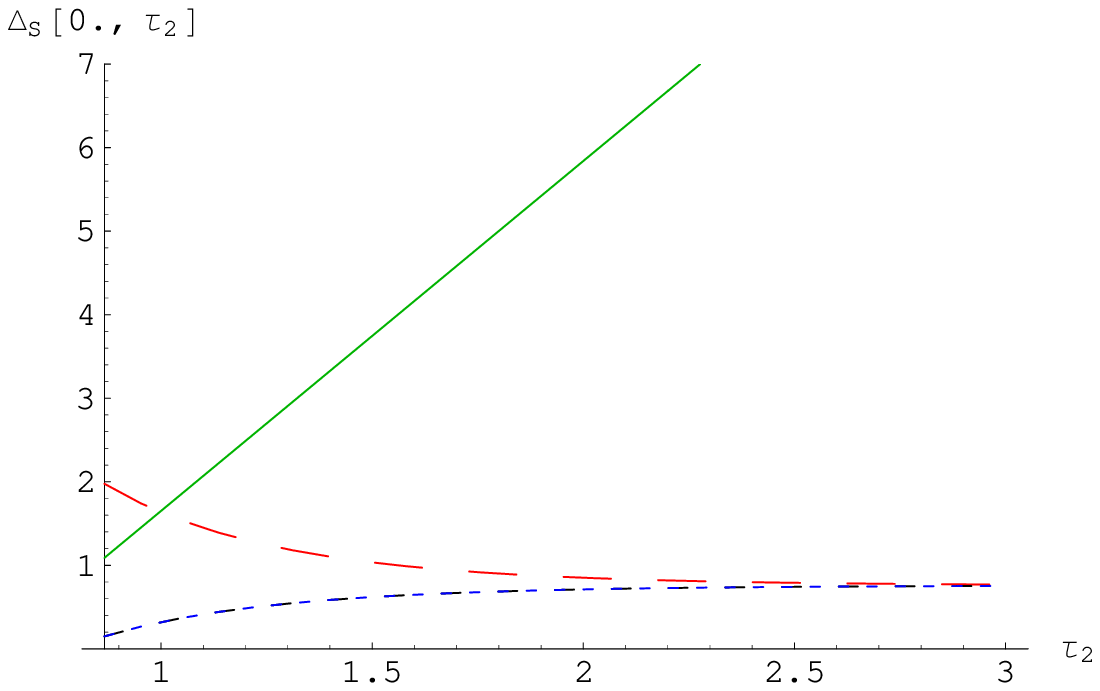} 
\\[2ex]
\includegraphics[width=3.7cm,height=3.2cm]{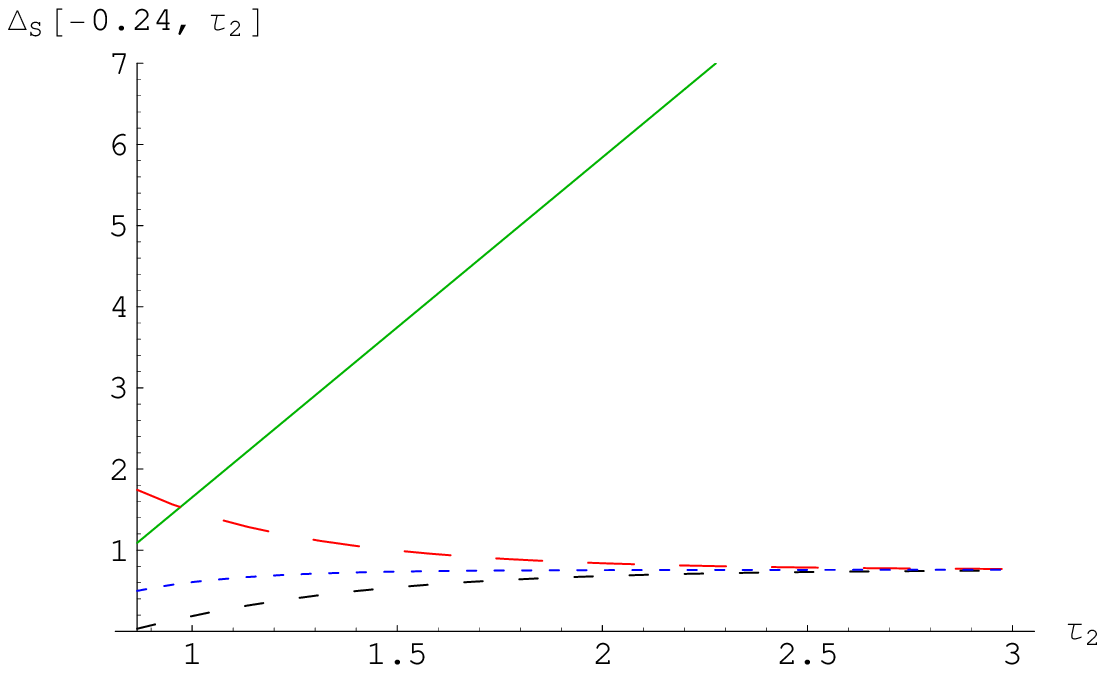} &
\includegraphics[width=3.7cm,height=3.2cm]{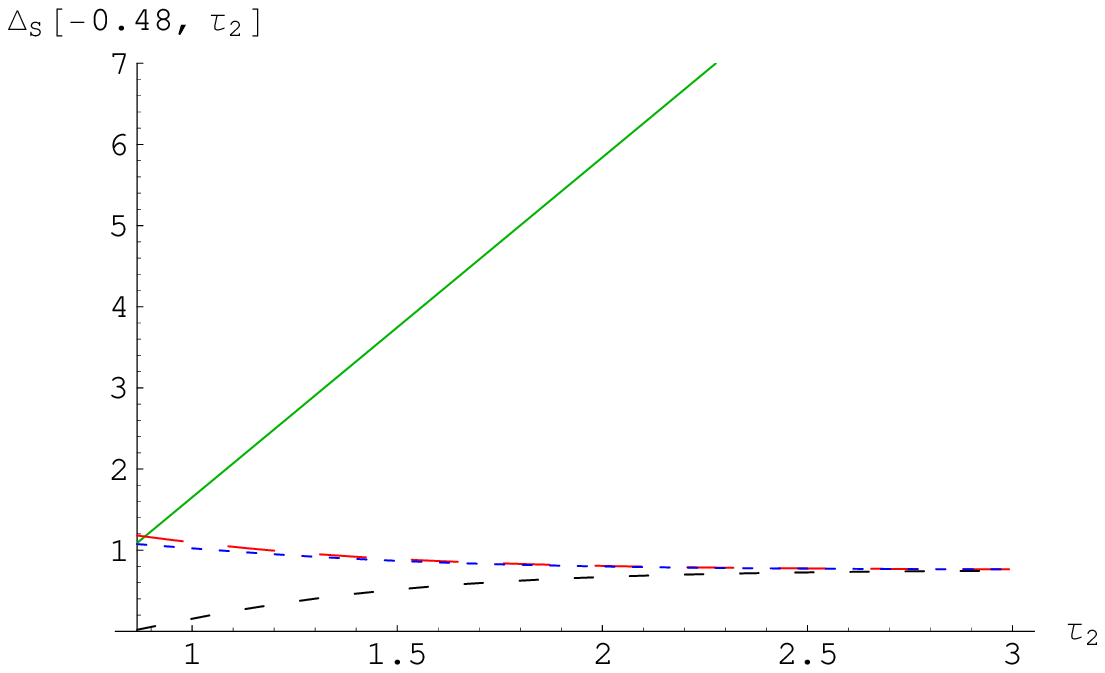} &
\raisebox{4ex}{\scalebox{0.5}{\mbox{\input{LngdgD.pstex_t}}}}
}
\end{center}
\caption{ 
The five plots show the differences of the functions 
$\gD_s(\gt_1,\gt_2)$ for the values $\gt_1 = -0.48, -0.24, 0, 0.24$ and
$0.48$ in the different sectors $s = u, t, d_+$ and $d_-$. (The values
were mainly motivated to show the changes in $\gD_s$ with $\gt_1$
clearly, and to avoid too many coincident curves.)  
The change of the function $\gD_u$ for these values of $\gt_1$ is
hardly  visible as can be seen from its approximation in equation
\eqref{ApprxgD}. The roles of $\gD_{d_+}$ and $\gD_{d_-}$ are
interchanged when we take  $\gt_1 \ra -\gt_1$. 
\labl{fg:gDplots}}
\end{figure}

\begin{table}[ht]
\tbl{Suppressed exponential factors.\labl{tb:ValuesExp}}{
\( 
\renewcommand{\arraystretch}{1.5}
\arry{| l | l l l l |}{
\hline 
\multicolumn{1}{|c}{\gt_2 } & \multicolumn{1}{|c}{\frac 12 \sqrt 3} 
& \multicolumn{1}{c}{1} & \multicolumn{1}{c}{1.5} 
&    \multicolumn{1}{c|}{2} 
\\ \hline 
~e^{- 2\pi \, \frac 13 \, \gt_2} ~ & 
~~0.163 & 0.123 & ~~0.043 & ~~0.015 
\\
~e^{- 2\pi \, \frac 23 \, \gt_2}  & 
~~0.027 & 0.015 & ~~0.002 & ~~0.000\, 2 
\\ 
~e^{- 2\pi \,\, \gt_2}  & 
~~0.004 & 0.002 & ~~0.000\, 1 & ~~0.000\, 03~
\\ \hline 
}
\) 
}
\end{table}

\noindent 
The full (on--shell) result for the gauge field tadpole in string
theory can now be written in a form that closely resembles the 
field theoretical expression given in \eqref{TadFieldSchwinger}: 
\equ{
\langle F_{j\uj}^{b}(k) \rangle = 
\frac{\gd^4(k_4)}{(2\gp)^4}\,  \sum_{s = u, t, d_\pm} 
\int_\cF  \frac {\d^2 \gt}{\gt_2^2}
\, Q_s^{b}(\gt) \, 
e^{- \gD_s(\gt,\bgt) \, k_i k_\ui }. 
\labl{SketchTadp}
}
To investigate the relationship  between the field and string
predictions for the gauge field tadpoles further, we expand the string
result as follows: As $\gt$ takes values within the fundamental domain
$\cF$, see figure \ref{fg:ModularPara}, the exponential factor
$\exp(-2 \pi \gt_2) < 1$ always suppressed. The relevant
exponentials for various small values of $\gt_2$ are given 
in table \ref{tb:ValuesExp}. Expanding the exact expressions for
$\gD_s(\gt,\bgt)$ to order $\exp(-2 \pi\gt_2)$ gives 
\equ{
\arry{lcl}{
\gD_u(\gt,\bgt) &= &
- 2 \ln 3 + 4 \gp  \gt_2 \frac {1}{3} + \ldots, 
\\[2ex]
\gD_t(\gt,\bgt) &= &
\ln 3  
+ 6 \cos(2 \gp\frac {\gt_1}3) e^{- 2 \pi \gt_2 \, \frac 13}
+ 9 \cos(4\gp \frac {\gt_1}3) e^{-2\pi \gt_2\,\frac 23}+ \ldots,
\\[2ex]
\gD_{d_+}(\gt,\bgt) &= &
\ln 3 
+ 6 \cos(2 \gp \frac {\gt_1-1}{3}) e^{- 2 \pi \gt_2\, \frac 13}
+ 9 \cos(4 \gp \frac {\gt_1-1}{3}) e^{- 2 \pi \gt_2\, \frac 23}
+ \ldots, 
\\[2ex]
\gD_{d_-}(\gt,\bgt) &=  &
\ln 3 
+ 6 \cos(2 \gp \frac {\gt_1+1}{3}) e^{- 2 \pi \gt_2\, \frac 13}
+ 9 \cos(4 \gp \frac {\gt_1+1}{3}) e^{- 2 \pi \gt_2\, \frac 23}
+ \ldots.
}
\labl{ApprxgD}
}
All these expressions are positive definite, as can inferred from
their plots given in figure \ref{fg:gDplots}. 
As discussion in ref.\ \cite{GNL_II} we have fixed the relative
normal ordering constant between the untwisted and twisted world sheet
bosons such that all $\gD_s$ are strictly positive. 
This is important as allows us to Fourier transform the
tadpole to coordinate space: 
\equ{
\langle F_{j \uj}^{b}(z) \rangle = 
\frac{\gd^4(k_4)}{(2\gp)^4}\,   
\sum_{s=u,t,d_\pm} 
\int_\cF  \frac {\d^2 \gt}{\gt_2^2}\, 
Q_s^{b}(\gt) \, 
\Bigl( \frac{1 }{ \gp\gD_s(\gt, \bgt)} \Bigr)^3
\,e^{-  \bar z z / \gD_s(\gt,\bgt)}. 
\labl{SketchTadpCoord}
}
The resulting profiles for the different sectors are plotted in figure
\ref{fg:SU9shapeCoor}. We see that the twisted sector contributions
$G_t$ and $G_{d_\pm}$ decay away very quickly for large $|z|$. 
Hence we see that the contributions of the twisted states are
localized within a couple of string lengths near the orbifold
singularity. The decay of the untwisted sector $G_u$ is much slower,
as has been expected since the untwisted sector corresponds to bulk --
non--localized -- contributions.

\begin{figure}
\begin{center}
\hspace{0cm}{
\tabu{ccc}{
\epsfig{file=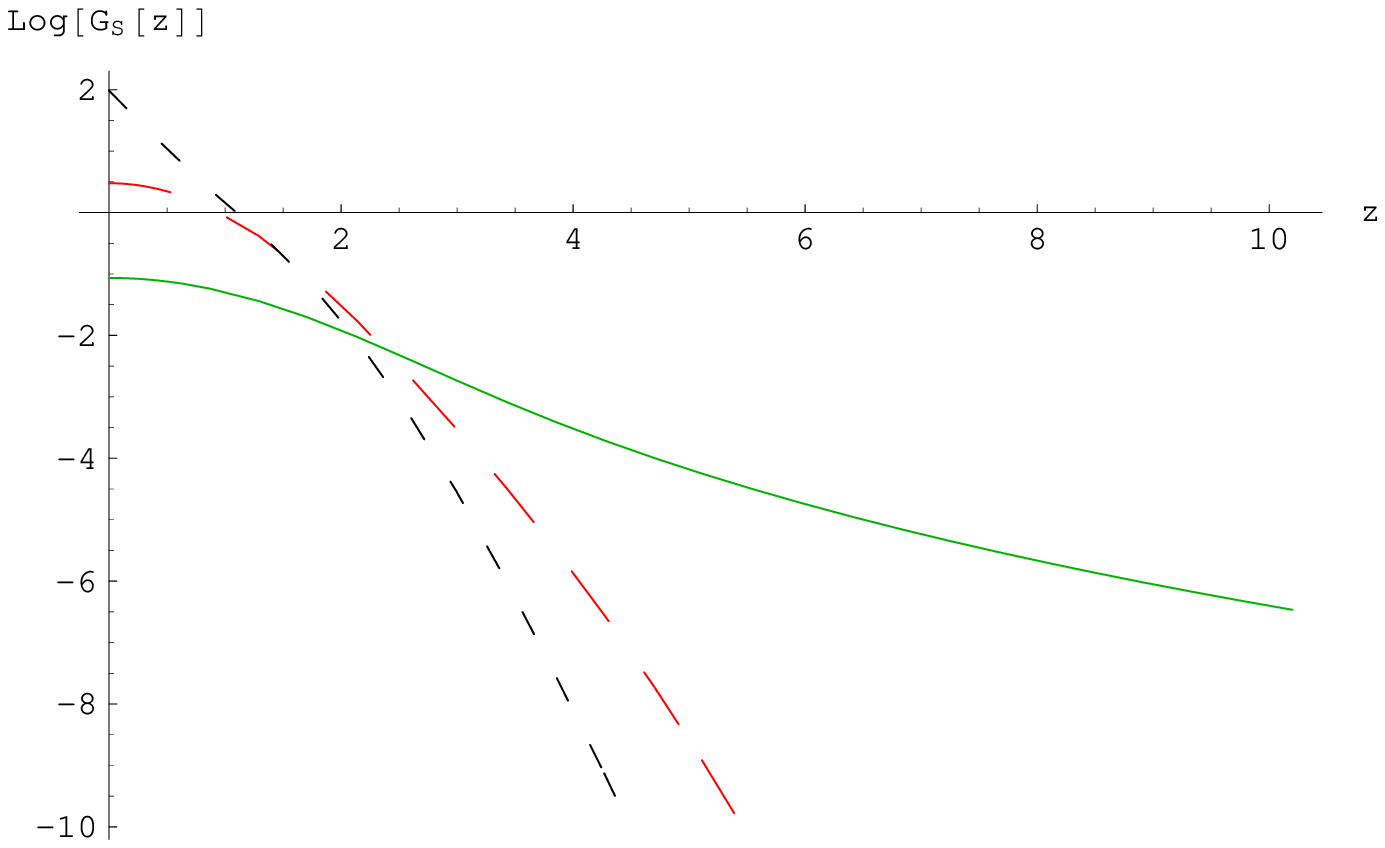,width=8cm}
& & 
\hspace{-1cm}{
\raisebox{4cm}{\scalebox{0.6}{\mbox{\input{LngdGauss.pstex_t}}}}}
}}
\end{center}
\caption{The spatial extension of the tadpole contributions in the
sectors $u, t$ and $d_\pm$ are displayed on a logarithmic scale.  
The curve for the $u$--sector fall off much slower than the others for
large values of $|z|$, as only it corresponds to bulk contributions.
\labl{fg:SU9shapeCoor}}
\end{figure}

\section{Local tachyons?}

In the discussion of the previous section we have (implicitly) assumed
that only the string zero modes contribute to the tadpoles. For the
two heterotic $\E{8}\times \E{8}$ models with anomalous $\U{1}$'s this
assumption is only true for the $\SU{9}$ model, given in table
\ref{tb:z3models}. For the other anomalous model, containing the
$\E{7}$ group, the situation is more complicated as can be seen from
the charge functions:  
\equ{
\arry{l}{
Q_u^1~~ \approx ~~ 3 + \ldots, 
\\[1ex] 
Q_t^1 ~~\approx ~~ 
\frac{1}{9}\,  q^{-\frac{1}{3}} + \frac{5}{3} + 6\,q^{\frac{1}{3}} - 
   \frac{76}{9} \,q^{\frac{2}{3}} + \ldots, 
\\[1ex]
Q_{d_+}^1 \approx 
~~ \frac{1}{9}\, e^{i \frac{4\pi}{3}} \,q^{-\frac{1}{3}} + \frac{5}{3} 
+ 6\,e^{i\frac{2\pi}{3}}\,q^{\frac{1}{3}} 
- \frac{76}{9}\, e^{i \frac{4 \pi}{3}} \, q^{\frac{2}{3}} + \ldots, 
\\[1ex]
Q_{d_-}^1 \approx 
~~  \frac{1}{9}\,e^{i\frac{2\pi}{3}}\, q^{-\frac{1}{3}} +  \frac{5}{3} 
+ 6\,e^{i \frac{4 \pi}{3}}\, q^{\frac{1}{3}} 
- \frac{76}{9}\, e^{i \frac{2\pi}{3}} \, q^{\frac{2}{3}} + \ldots, 
}
\labl{Qexp1}
}
and 
\equ{
\arry{l}{
Q_u^2 ~~ \approx ~~  2 + \ldots,  
\\[1ex]
Q_t^2 ~~ \approx ~~ 
  \frac{2}{9}\,q^{-\frac{1}{3}} - \frac{2}{3} + 12\,q^{\frac{1}{3}} - 
   \frac{152}{9} \,q^{\frac{2}{3}} + \ldots, 
\\[1ex] 
Q_{d_+}^2 \approx 
 ~~ \frac{2}{9} \,e^{i \frac{4\pi}{3}} \,q^{-\frac{1}{3}} -  \frac{2}{3} 
+ 12\, e^{i \frac{2\pi}{3}}\, q^{\frac{1}{3}} 
- \frac{152}{9}\, e^{i\frac{4 \pi}{3}}\, q^{\frac{2}{3}} + \ldots, 
\\[1ex]
Q_{d_-}^2 \approx 
~~ \frac{2}{9} \, e^{i\frac{2\pi}{3}} \, q^{-\frac{1}{3}} - \frac{2}{3}  
+ 12\, e^{i \frac{4 \pi}{3}}\, q^{\frac{1}{3}} 
- \frac{152}{9}\, e^{i \frac{2\pi}{3}}\, q^{\frac{2}{3}} + \ldots. 
}
\labl{Qexp2}
} 
The negative power $q^{- 1/3}$ indicate that tachyonic states
contribute to the charge functions $Q^b_s$ for the twisted sectors 
$s = t, d_+$ and $d_-$. (The positive powers result from massive string
excitations. In fact, a tower of massive string states contribute to
these local shapes.)

The presence of the tachyonic contributions is worrying, since it
tachyons are normally taken as an indication of instabilities. But since
the orbifold is supersymmetric it cannot be unstable. To understand
what is going on, we consider the expression of the tachyonic
contributions only in the coordinates space representation: 
\equ{
\cT(z) = 
\int_\cF  \frac {\d^2 \gt}{\gt_2^2} 
\, e^{2 \pi \, \frac {\gt_2}3 }
\left\{ 
\cos \bigl( \frac {2\pi}3 \gt_1 \bigr) 
\Bigl( 
\frac{e^{- |z|^2/  \gD_t(\gt) } }{ (\gp \gD_t(\gt) )^3 } 
- 
\cE_+(k| \gt) \Bigr) 
- \sqrt 3\, \sin \bigl( \frac {2\pi}3 \gt_1 \bigr) \cE_-(k | \gt) 
\right\}. 
\labl{CoorTachTadpReal}
}
Like the exponential factor, the even and odd functions 
\equ{
\cE_\pm(z|\gt) = \frac 12 
\Bigl( 
\frac{ e^{- |z|^2/ \gD_{d_+}(\gt)} }{(\gp \gD_{d_+}(\gt))^3 } 
\pm 
\frac{ e^{- |z|^2/ \gD_{d_-}(\gt)} }{(\gp \gD_{d_-}(\gt))^3} 
\Bigr), 
\qquad 
\cE_\pm(z| -\gt_1, \gt_2) = \pm \cE_\pm(z| \gt_1, \gt_2),
} 
contain the six dimensional Gaussian normalization factors 
$1/(\gp \gD_{d_\pm})^3$. Therefore, the integral over the orbifold of
each of these exponentials is normalized to unity. It follows that the
integrated tadpole due to tachyons vanish. This implies the
cancellation of the tachyonic contributions within the zero mode
theory, which confirms that the model is globally stable.

\begin{figure}
\begin{center}
\epsfig{file=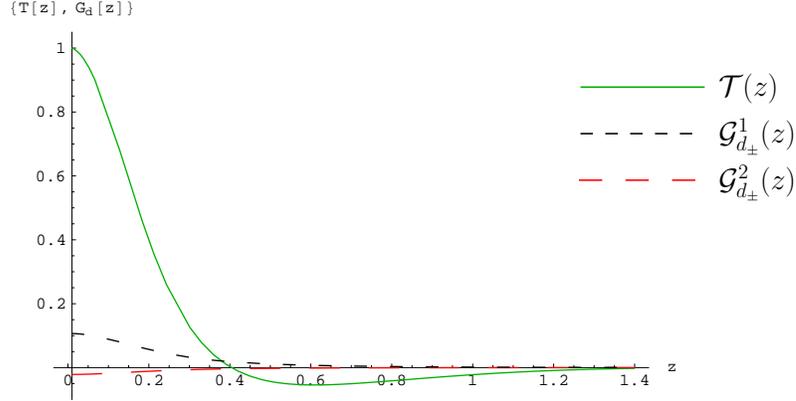,width=9cm}
\put(-40,80){\scalebox{0.65}{\mbox{\input{LngdTachZero.pstex_t}}}}
\end{center}
\caption{
The coordinate space representation of the tachyonic contributions to
the gauge field tadpole is displayed by the solid line. This
contribution  dominates the largest zero mode contribution
coming from the $d_\pm$ sectors. 
\labl{fg:CoorTach}}
\end{figure}

In figure \ref{fg:CoorTach} the tachyonic contribution to the tadpole
is plotted together with the leading contributions from the zero
modes. Inspecting figure  \ref{fg:SU9shapeCoor} we infer that the
twisted sectors $d_{\pm}$ give the largest contributions to the zero
mode states. (The other sectors contributions are at least one
order of magnitude less, and can therefore safely be ignored in the
present analysis.) The normalization of the zero modes in
the $d_\pm$ sectors compared to the tachyonic sector is $15$ and $-3$
for the first and second $E_8$, respectively, see equations
\eqref{Qexp1} and \eqref{Qexp2}. These relative
normalizations have been taken into account in comparison figure 
\ref{fg:CoorTach}. We conclude that the tachyonic states
totally dominate  the profile of the tadpoles for the $\U{1}$'s in
both $\E{8}$'s near the singularity.

This curious appearance of local tachyonic contributions to gauge
field tadpoles does not only appear in the heterotic $\E{8}\times
\E{8}$ theory, but is also present in some $\SO{32}$ string
models. There are five heterotic $\SO{32}$ models $\Intr_3$ classified
according to their gauge shift vector 
$v= (1^{2n}, -2^n, 0^{16-3n})/3$, $n=1,\ldots 5$, with an anomalous
$\U{1}$. 
In the $n=2$ and $5$ models local tachyons are present as well, while
in the others, like the $\SU{9}$ model, only zero modes contribute to
the tadpole. The full implications of these local tachyons have not
been uncovered yet and might provide an interesting avenue for future
research.

\section{Conclusions}

We began this review by considering field theories on orbifolds. Even
though such theories have many useful and interesting properties, the
physics near the orbifold singularity is beyond the realm of field
theory. To investigate just how far one can push the field theory
analysis we compute gauge field tadpoles on a $\Intr_3$ orbifold and
found singular behavior near the orbifold fixed point where the field
theory calculation might not be trusted. We considered heterotic
string models as UV completions of the field theory and computed the
gauge field tadpoles in this framework as well. Away from the orbifold
singularity the qualitative properties of both computations are the
same. Near the orbifold singularity things are different. In
particular, the twisted states which in field theory are assumed to be
localized exactly at the orbifold fixed point, have in string theory a
finite extend of a couple of string lengths. Moreover, contrary to the
field theory result, string theory gives strictly finite answer.

These string results could have been anticipated, what has been a
major surprise of our computation of the gauge field tadpoles in
string theory is that for some orbifold models there are, aside from
the zero modes, contributions from local tachyonic states to these
amplitudes. Yet, these tachyons do not signify a global instability
since they are not present in the spectrum of the theory, contributing
only in the loop. Moreover, when integrated over the whole orbifold,
these tachyonic contributions precisely cancel each other.


\section*{Acknowledgments}

This work was partially supported by DOE grant DE--FG02--94ER--40823
at the University of Minnesota.

\section*{References}

\bibliographystyle{ws-mpla}
\bibliography{paper}

\providecommand{\href}[2]{#2}\begingroup\raggedright\begin{thebibliography}{10}

\bibitem{Fayet:1974jb}
P.~Fayet and J.~Iliopoulos {\em Phys. Lett.} {\bf B51}, 461--464
(1974).

\bibitem{Fischler:1981zk}
W.~Fischler, H.~P. Nilles, J.~Polchinski, S.~Raby, and L.~Susskind {\em Phys.
  Rev. Lett.} {\bf 47}, 757
(1981).

\bibitem{Dine:1987xk}
M.~Dine, N.~Seiberg, and E.~Witten {\em Nucl. Phys.} {\bf B289}, 589
(1987).

\bibitem{Atick:1987gy}
J.~J. Atick, L.~J. Dixon, and A.~Sen {\em Nucl. Phys.} {\bf B292}, 109--149
(1987).

\bibitem{Dine:1987gj}
M.~Dine, I.~Ichinose, and N.~Seiberg {\em Nucl. Phys.} {\bf B293}, 253
(1987).

\bibitem{Poppitz:1998dj}
E.~Poppitz {\em Nucl. Phys.} {\bf B542}, 31--44 (1999)
\href{http://www.arXiv.org/abs/hep-th/9810010}{[{\tt hep-th/9810010}]}.

\bibitem{Mirabelli:1998aj}
E.~A. Mirabelli and M.~E. Peskin {\em Phys. Rev.} {\bf D58}, 065002 (1998)
\href{http://www.arXiv.org/abs/hep-th/9712214}{[{\tt hep-th/9712214}]}.

\bibitem{Ghilencea:2001bw}
D.~M. Ghilencea, S.~Groot~Nibbelink, and H.~P. Nilles {\em Nucl. Phys.} {\bf
  B619}, 385--395 (2001)
\href{http://www.arXiv.org/abs/hep-th/0108184}{[{\tt hep-th/0108184}]}.

\bibitem{Barbieri:2001cz}
R.~Barbieri, L.~J. Hall, and Y.~Nomura
\href{http://www.arXiv.org/abs/hep-ph/0110102}{[{\tt hep-ph/0110102}]}.

\bibitem{Scrucca:2001eb}
C.~A. Scrucca, M.~Serone, L.~Silvestrini, and F.~Zwirner {\em Phys. Lett.} {\bf
  B525}, 169--174 (2002)
\href{http://www.arXiv.org/abs/hep-th/0110073}{[{\tt hep-th/0110073}]}.

\bibitem{GrootNibbelink:2002wv}
S.~Groot~Nibbelink, H.~P. Nilles, and M.~Olechowski {\em Phys. Lett.} {\bf
  B536}, 270--276 (2002)
\href{http://www.arXiv.org/abs/hep-th/0203055}{[{\tt hep-th/0203055}]}.

\bibitem{GrootNibbelink:2002qp}
S.~Groot~Nibbelink, H.~P. Nilles, and M.~Olechowski {\em Nucl. Phys.} {\bf
  B640}, 171--201 (2002)
\href{http://www.arXiv.org/abs/hep-th/0205012}{[{\tt hep-th/0205012}]}.

\bibitem{Lee:2003mc}
H.~M. Lee, H.~P. Nilles, and M.~Zucker {\em Nucl. Phys.} {\bf B680}, 177--198
  (2004)
\href{http://www.arXiv.org/abs/hep-th/0309195}{[{\tt hep-th/0309195}]}.

\bibitem{Barbieri:2002ic}
R.~Barbieri, R.~Contino, P.~Creminelli, R.~Rattazzi, and C.~A. Scrucca {\em
  Phys. Rev.} {\bf D66}, 024025 (2002)
\href{http://www.arXiv.org/abs/hep-th/0203039}{[{\tt hep-th/0203039}]}.

\bibitem{Marti:2002ar}
D.~Marti and A.~Pomarol {\em Phys. Rev.} {\bf D66}, 125005 (2002)
\href{http://www.arXiv.org/abs/hep-ph/0205034}{[{\tt hep-ph/0205034}]}.

\bibitem{vonGersdorff:2002us}
G.~von Gersdorff, N.~Irges, and M.~Quiros {\em Phys. Lett.} {\bf B551},
  351--359 (2003)
\href{http://www.arXiv.org/abs/hep-ph/0210134}{[{\tt hep-ph/0210134}]}.

\bibitem{Csaki:2002ur}
C.~Csaki, C.~Grojean, and H.~Murayama {\em Phys. Rev.} {\bf D67}, 085012 (2003)
\href{http://www.arXiv.org/abs/hep-ph/0210133}{[{\tt hep-ph/0210133}]}.

\bibitem{GrootNibbelink:2003gb}
S.~Groot~Nibbelink, H.~P. Nilles, M.~Olechowski, and M.~G.~A. Walter {\em Nucl.
  Phys.} {\bf B665}, 236--272 (2003)
\href{http://www.arXiv.org/abs/hep-th/0303101}{[{\tt hep-th/0303101}]}.

\bibitem{GNL_I}
S.~Groot~Nibbelink and M.~Laidlaw {\em JHEP} {\bf 01}, 004 (2004)
\href{http://www.arXiv.org/abs/hep-th/0311013}{[{\tt hep-th/0311013}]}.

\bibitem{Polchinski:1986zf}
J.~Polchinski {\em Commun. Math. Phys.} {\bf 104}, 37
(1986).

\bibitem{GNL_II}
S.~Groot~Nibbelink and M.~Laidlaw {\em JHEP} {\bf 01}, 036 (2004)
\href{http://www.arXiv.org/abs/hep-th/0311015}{[{\tt hep-th/0311015}]}.

\bibitem{Antoniadis:2002cs}
I.~Antoniadis, E.~Kiritsis, and J.~Rizos {\em Nucl. Phys.} {\bf B637}, 92--118
  (2002)
\href{http://www.arXiv.org/abs/hep-th/0204153}{[{\tt hep-th/0204153}]}.

\end{thebibliography}\endgroup

\end{document}